\documentclass[sigconf,natbib=true,anonymous=false,nonacm]{acmart}
\AtBeginDocument{%
  }

\usepackage{multirow}
\usepackage{array} 
\usepackage{color}
\usepackage{soul}
\DeclareRobustCommand{\hlyellow}[1]{{\sethlcolor{yellow!20}\hl{#1}}}
\DeclareRobustCommand{\hlgreen}[1]{{\sethlcolor{green!10}\hl{#1}}}

\usepackage{tikz}

\newcommand{\figlabeld}[1]{%
    \protect\tikz[baseline=(X.base)] \protect\node[draw, 
    rounded corners=2pt, inner sep=1.5pt, font=\footnotesize\sffamily, text height=1.5ex, text depth=0.25ex] (X) {#1};\hspace{1pt}%
}
\newcommand{\figlabeln}[1]{%
    \protect\tikz[baseline=(X.base)] \protect\node[draw, 
    rounded corners=2pt, inner sep=1.5pt, font=\footnotesize\sffamily, text height=1.5ex, text depth=0.25ex] (X) {#1};\hspace{1pt}%
}

\begin{document}

\title{NumColBERT: Non-Intrusive Numeracy Injection for Late-Interaction Retrieval Models}

\author{Haruki Fujimaki}
\affiliation{%
  \institution{University of Tsukuba}
  \city{Tsukuba}
  \country{Japan}
}
\email{fujimaki.haruki.tkb\_eg@u.tsukuba.ac.jp}
\orcid{0009−0000−2209−7171}

\author{Makoto P. Kato}
\affiliation{%
  \institution{University of Tsukuba}
  \city{Tsukuba}
  \country{Japan}}
\affiliation{%
  \institution{National Institute of Informatics}
  \city{Tokyo}
  \country{Japan}}
\email{mpkato@acm.org}
\orcid{0000−0002−9351−0901}

\renewcommand{\shortauthors}{H.Fujimaki and M.P.Kato}

\begin{abstract}
This study addresses the challenge of improving dense retrieval performance for queries containing numerical conditions, such as ``companies with more than one billion dollars in R\&D expenditure.'' Although recent research has underscored the limitations of standard models in handling numeric information across domains such as finance, e-commerce, and medicine, existing solutions typically decompose queries into textual and numerical components and score them separately using dedicated methods. These approaches intrude upon late-interaction retrieval models such as ColBERT and incur considerable challenges in deployment, latency, and maintainability.
To overcome these limitations, we propose NumColBERT, an \textit{inference-time non-intrusive} method that enhances numerically conditioned retrieval while preserving the original late-interaction mechanism and providing unified scoring across textual and numerical content. Because NumColBERT retains the standard ColBERT indexing and MaxSim scoring pipeline, existing optimizations and ecosystem components developed for ColBERT can be directly reused, facilitating practical deployment.
NumColBERT introduces a Numerical Gating Mechanism and a Numerical Contrastive Learning objective to enable numerical conditions to contribute more effectively to retrieval within the standard ColBERT scoring mechanism. The gating mechanism dynamically amplifies the influence of tokens carrying critical numerical constraints while suppressing context-neutral mentions such as model numbers or dates. The contrastive objective explicitly shapes the embedding space to reflect numerical magnitudes and conditions, enabling numerical values to be distinguished within the shared representation space.
Experimental results show that NumColBERT substantially outperforms standard fine-tuning baselines and achieves accuracy that matches or exceeds that of prior approaches that rely on separate textual and numerical scoring. These findings demonstrate the feasibility of numerically conditioned retrieval with a non-intrusive inference pipeline and present a maintainable solution for real-world deployment.

\end{abstract}




\maketitle

\section{Introduction}

In recent years, dense retrieval models have been widely adopted across a diverse range of domains, including product search, financial document retrieval, and medical literature search~\cite{reddy2022shoppingqueriesdatasetlargescale,gill2025keywords,chen2022finqadatasetnumericalreasoning,jin2020diseasedoespatienthave,mahendra-etal-2024-numbers}.
Dense retrieval models encode queries and documents into dense vectors using language models such as BERT~\cite{devlin-etal-2019-bert}, and perform retrieval based on similarity scores between these vectors. Compared with traditional lexical matching approaches, dense retrieval enables context-aware search that considers semantic information contained in queries and documents.

On the other hand, it has been repeatedly noted that language models exhibit insufficient understanding of numerical information~\cite{wallace-etal-2019-nlp,dua2019dropreadingcomprehensionbenchmark}.
Consequently, dense retrieval models built upon such language models are likewise expected to struggle when handling queries containing numerical constraints.
Indeed, Fujimaki et al.~\cite{Haruki2025investigating-numqueries} and Almasian et al.~\cite{almasian2024numbersmatterbringingquantityawareness-numbersmatter} analyzed the performance of dense retrieval models on numerically conditioned queries and revealed a substantial gap between expected and actual performance. Their findings in particular highlight limited capability in numeric comparison within dense retrieval models.

This limitation raises concerns regarding the reliability of information access in quantity-aware retrieval scenarios.
Typical examples include retrieving ``disasters reported to have caused more than 1,000 fatalities'' from news articles, or searching clinical reports for ``patients whose blood pressure decreased by more than 20 mmHg after treatment.''
Such information needs require the correct interpretation of numbers, units, and comparison operators embedded in unstructured natural language and prioritizing documents that satisfy the requested conditions.
These requirements cannot be fully addressed by structured database queries such as SQL, making it essential for dense retrieval models operating over unstructured text to acquire the ability to handle numerical conditions.

Existing approaches aimed at improving numerically conditioned retrieval include the work of Almasian et al.~\cite{almasian2024numbersmatterbringingquantityawareness-numbersmatter} and Agrawal et al.~\cite{agrawal-etal-2025-dense-deepquant}.
These studies achieve high retrieval performance by integrating components such as quantity and unit extraction modules, neural networks specialized for numeric comparison, or scoring mechanisms designed for compatible unit systems.
However, these methods require additional specialized processing outside the retrieval model or introduce dedicated numerical scoring beyond standard embedding similarity during inference, necessitating modifications to the retrieval process and leaving challenges regarding integration, latency, and operational cost.

In this work, we aim to explore to what extent numerically conditioned retrieval performance can be improved without introducing specialized external modules or changing inference logic, under the constraint of preserving the existing ColBERT~\cite{10.1145/3397271.3401075colbert,santhanam-etal-2022-colbertv2} search pipeline.
In this study, {\it preserving the retrieval pipeline} means making no modifications to core components such as document indexing or similarity computation (MaxSim), and performing all processing related to numerical conditions entirely within the token embedding generation process.
Specifically, we introduce embedding-space enhancements and auxiliary training tasks designed to encourage numerical understanding only during training, while retaining the standard ColBERT inference configuration that relies solely on token embeddings.
Our approach comprises two core innovations.
First, the \textit{Numerical Gating Mechanism} addresses the structural limitation that numerical tokens contribute minimally to the MaxSim score when other tokens exhibit strong semantic matches~\cite{hofstatter2022colberter,kang-etal-2025-trial}; by applying learned scaling factors to numerical tokens, the gate amplifies their contribution to the similarity computation while preserving compatibility with standard ColBERT indexing and inference.
Second, the \textit{Numerical Contrastive Loss} explicitly shapes the embedding space for numerical reasoning; during training, documents satisfying the query's numerical condition are brought closer in the embedding space while those violating the condition are separated, encouraging the encoder to capture condition satisfaction within the standard similarity-based retrieval framework.

Experimental results on benchmark datasets in the finance and medical domains demonstrate that the proposed approach achieves significant improvements over fine-tuning of ColBERT and provides performance comparable to or better than prior work incorporating numerical-specialized modules.
This paper shows that, even under strict constraints that prohibit altering the retrieval pipeline, numerical reasoning capabilities for constrained queries can be enhanced solely through training strategies and embedding design. We highlight this direction as a promising path toward practical, low-cost, and high-precision numerically conditioned retrieval.

The contributions of this work are summarized as follows.
\begin{itemize}
    \item We propose \textbf{NumColBERT}, a late-interaction model for numerically conditioned retrieval.
    It is non-intrusive at inference time and fully compatible with ColBERT.
    \item We empirically demonstrate that NumColBERT achieves accuracy equivalent to or exceeding specialized-module approaches on numerically conditioned retrieval tasks in finance and medical domains.
    \item We establish the feasibility of numerical reasoning under pipeline-preservation constraints, offering new insights for the design of practical quantity-aware retrieval systems.
\end{itemize}

\section{Related Work}

\subsection{Language Models and Numeracy}
While Pre-trained Language Models (PLMs) have achieved remarkable success in general NLP tasks, their ability to reason about quantities remains a subject of active research.
Standard PLMs treat numbers as ordinary tokens, often failing to capture their inherent magnitude and continuity~\cite{geva2020injecting,yang-etal-2025-numbercookbook}.
To address this, several works have proposed enhancing numeracy in PLMs through specialized pre-training objectives~\cite{geva2020injecting, chen2023improving},
numerical-aware embeddings such as scientific notation or digit-level embeddings~\cite{sundararaman2020methods, thawani2021representing, sivakumar2025leverage},
or semantic priming strategies~\cite{sharma2024laying}.
However, these advancements primarily focus on improving the internal representation of numbers for generation or question-answering tasks.
They do not directly address the challenges of \textit{retrieving} documents from a large corpus based on numerical constraints, which requires distinct scoring mechanisms and index structures.

\subsection{Quantity-Aware Retrieval}
Early attempts to bridge the gap between text retrieval and numerical queries largely relied on extraction-based pipelines.
Systems like QSearch~\cite{ho2019qsearch} and others~\cite{ho2020entities, ho2021extracting, ho2021QuTE, ho2022enhancing} extract quantities (value and unit) from text and construct a structured knowledge base to enable SQL-like querying.
Similarly, AnaSearch~\cite{li2021anasearch} processes unstructured text to answer analytical queries.
While effective for structured data, these approaches suffer from the rigidity of extraction pipelines and lack the flexibility of ad-hoc retrieval.
Other works have explored reasoning about quantities in QA~\cite{roy2015reasoning} or web tables~\cite{ibrahim2016making}, but they are often limited to specific domains or tabular data~\cite{rybinski2023sciharvester} and do not scale well to open-domain dense retrieval.
Importantly, these extraction-based systems are designed for structured querying over pre-built knowledge bases, which fundamentally differs from the ad-hoc dense retrieval setting addressed in this work.

\subsection{Dense Retrieval with Quantity Comparison}
Recent work has begun to address dense retrieval for numerically conditioned queries, where relevance depends on satisfying comparison constraints such as ``more than 1B USD.''
One representative direction is the work by \citet{almasian2024numbersmatterbringingquantityawareness-numbersmatter}, which demonstrates strong performance with a \textit{disjoint} design---that is, textual and numerical information are processed through separate pathways, and the resulting signals are combined only at the final scoring stage.
This line of work relies on identifying values, units, and comparison operators using tools such as the Comprehensive Quantity Extractor (CQE)~\cite{almasian-etal-2023-cqe}, which parses free text to locate numerical mentions, normalize values, and attach semantic units and comparison operators.
The extracted numerical information is then handled in a dedicated pathway whose signal is integrated with the textual relevance.
Although effective, this design introduces an additional subsystem grounded in extraction and separate numeric processing, which increases pipeline complexity and operational cost.

DeepQuant~\cite{agrawal-etal-2025-dense-deepquant} makes the same assumption: quantities should be explicitly extracted by CQE at the input stage.
However, it differs in where the separation is enforced: instead of maintaining a separate index and retrieval path, DeepQuant computes two scores during late interaction.
A standard ColBERT-style textual score is computed over non-numeric tokens, while a numeric score is produced by a dedicated comparison module operating over the quantities extracted by CQE.
The two scores are then combined by a learned, query-dependent mixing weight.
This scheme retains strong performance while still realizing numerical reasoning through two distinct scoring channels that sit alongside one another rather than within a single unified late-interaction computation.

In contrast, NumColBERT, our proposed model, does not rely on CQE or any external extraction step at inference time.
Instead of creating a separate numeric pathway or score, numerical tokens are processed in the same embedding space, and their contribution is modulated directly within MaxSim.
As a result, NumColBERT preserves ColBERT's original indexing and scoring pipeline, achieving numerical conditioning in a unified manner without subsystem additions.

\subsection{Token-Level Gating in Late-Interaction Retrieval}
Recent work has explored learned gating as a way to control token contributions in neural retrieval models.
\citet{hofstatter2020learning} introduced contextualized stopwords, where token importance values are learned rather than fixed, showing that low-utility tokens can be suppressed based on context.
The idea was extended to late interaction by \citet{hofstatter2022colberter} with ColBERTer, which applies gating to drop low-value document tokens during indexing, reducing storage and inference cost while preserving effectiveness.
\citet{kang-etal-2025-trial} further showed that continuous gating can improve retrieval quality by upweighting useful tokens rather than only removing uninformative ones, demonstrating that gating can enhance ranking effectiveness, not just efficiency.

These prior gating methods target either efficiency (by pruning low-utility tokens) or general importance weighting (by modulating all tokens uniformly).
In contrast, NumColBERT introduces a \textit{Numerical Gating Mechanism} that specifically targets numerical tokens and modulates their contribution according to comparison semantics.
Rather than learning a universal importance score, our gate is conditioned on the magnitude and comparison intent of numerical values, enabling fine-grained control over how numerical constraints influence retrieval.


\section{Methodology}

In this section, we first formalize the problem of ad-hoc retrieval with numerical conditions and revisit the standard ColBERT architecture. We then analyze the limitations of existing state-of-the-art approaches, specifically DeepQuant. Finally, we introduce \textbf{NumColBERT}, our proposed method that enhances numerical sensitivity within the standard late-interaction retrieval pipeline through a gating mechanism and multi-task learning objectives.

\begin{figure*}[t]
    \centering
    \includegraphics[width=\linewidth]{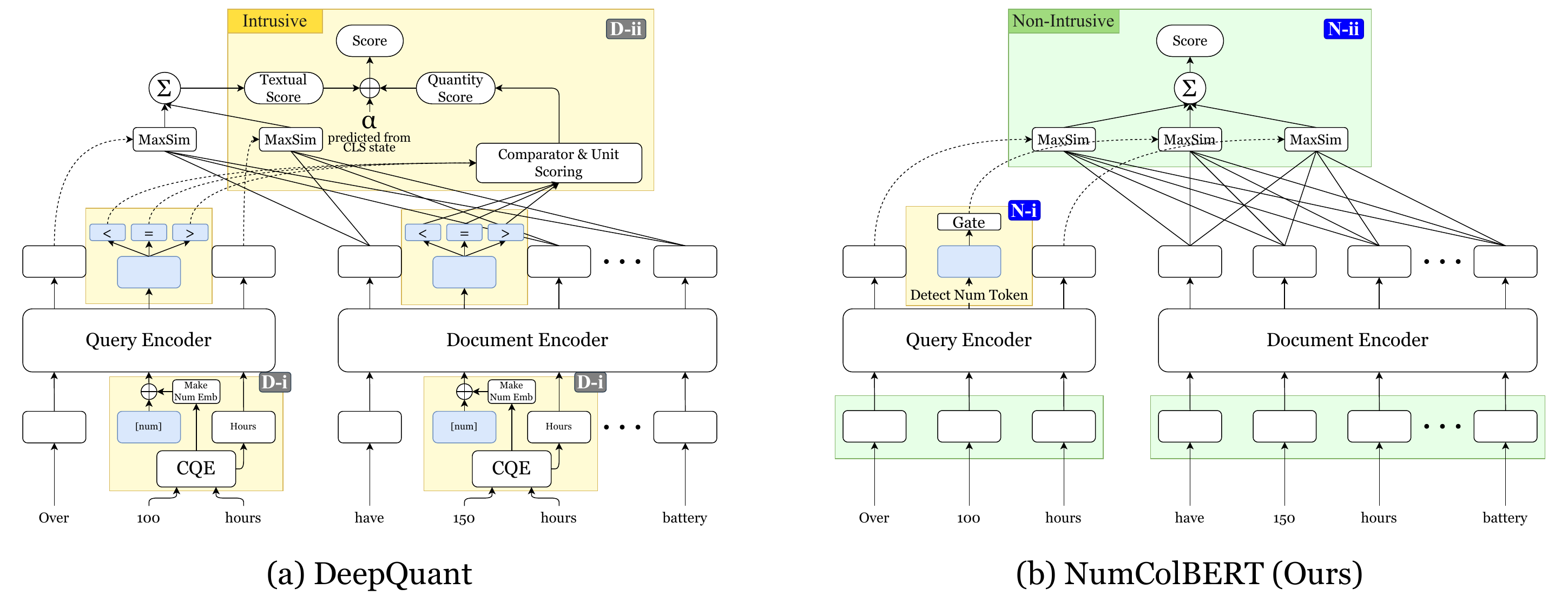}
    \caption{
    Architecture comparison between DeepQuant and NumColBERT. \hlyellow{Yellow-highlighted regions} indicate modifications from the standard ColBERT architecture, while \hlgreen{green-highlighted regions} indicate components unchanged from ColBERT. DeepQuant (left) consists of \figlabeld{D-i}~an external Comprehensive Quantity Extractor (CQE) for pre-processing and \figlabeld{D-ii}~a disjoint scoring pipeline that separately computes text and numerical scores. In contrast, NumColBERT (right) introduces \figlabeln{N-i}~a lightweight Numerical Gate with Numerical Token Detector applied only to query embeddings, while \figlabeln{N-ii}~the retrieval scoring remains the standard ColBERT MaxSim, preserving full compatibility with existing infrastructure.}
    \Description{Side-by-side architectural diagram comparing DeepQuant and NumColBERT. DeepQuant uses an external CQE module and a disjoint scoring pipeline combining separate text and numerical scores. NumColBERT applies a lightweight numerical gate only on the query side and keeps the standard ColBERT MaxSim scoring on the document side.}
    \label{fig:method_overall}
\end{figure*}

\subsection{Problem Formulation and Preliminaries}

We address the task of ad-hoc information retrieval where a query $Q$ may contain \textit{numerical conditions}. Formally, a numerical condition is a tuple $(v, \mathit{cmp}, u)$ specifying a numerical threshold $v$, a comparison operator $\mathit{cmp} \in \{=, <, > \}$, and an optional unit $u$ (e.g., ``USD'' and ``\%''). A document $D$ \textit{satisfies} the condition if it contains a quantity $(v', u')$ such that the comparison $v' \; \mathit{cmp} \; v$ holds and units are compatible. 
For instance, the query ``SSDs with capacity over 500 GB'' contains the condition $(500, >, \text{GB})$: ``1 TB'' (= 1,000 GB) satisfies this condition through unit conversion, while ``256 GB'' fails due to value mismatch. However, ``600 Mbps'' cannot be evaluated as transfer speed and storage capacity are dimensionally incompatible.

Standard \textbf{ColBERT} encodes queries and documents into sets of token embeddings.
Let $Q = \{\mathbf{q}_1, \ldots, \mathbf{q}_m\}$ denote the set of $m$ query token embeddings, and $D = \{\mathbf{d}_1, \ldots, \mathbf{d}_n\}$ the set of $n$ document token embeddings.
The ColBERT retrieval score is defined as the sum of maximum similarities (MaxSim):
\begin{equation}
    S(Q, D) = \sum_{\mathbf{q}_i \in Q} \max_{\mathbf{d}_j \in D} \mathbf{q}_i \cdot \mathbf{d}_j
\end{equation}
While ColBERT excels at semantic matching, it faces two fundamental limitations when handling numerical conditions:
\begin{enumerate}
    \item \textbf{Representation-level limitation.} The subword tokenization and dense encoding do not explicitly model magnitude, order, or operator semantics; hence, embeddings for ``2,000'' and ``20,000'' may be nearly indistinguishable.
    \item \textbf{Scoring-structure limitation.} Under MaxSim, each token contributes at most once. Since a numerical condition typically spans only 1--2 tokens while the query may contain many lexical terms, the numerical contribution is easily overshadowed by strong semantic matches elsewhere.
\end{enumerate}
As a result, even when a document violates a critical numerical constraint, high lexical overlap can ``push through'' the score, causing ranking failures.

\subsection{Limitations of Prior Work}

Recent state-of-the-art approaches, such as DeepQuant~\cite{agrawal-etal-2025-dense-deepquant}, address these limitations by explicitly separating numerical processing from semantic matching.
DeepQuant employs an external Comprehensive Quantity Extractor (CQE)~\cite{almasian-etal-2023-cqe} module to decompose queries and documents into text and quantity sets: $Q = \{Q_{\text{text}}, Q_{\text{num}}\}$ and $D = \{D_{\text{text}}, D_{\text{num}}\}$.
The final scoring function is a weighted sum of two distinct scores:
\begin{align}
    S_{\text{DeepQuant}}(Q, D) &= (1-\alpha_Q) \cdot S_{\text{num}}(Q_{\text{num}}, D_{\text{num}}) \notag \\
    &\quad + \alpha_Q \cdot S_{\text{text}}(Q_{\text{text}}, D_{\text{text}})
\end{align}
Here, $S_{\text{num}}$ is computed via a specialized network that marginalizes over \textit{comparison operator} uncertainty (i.e., which comparison operator among $=$, $<$, and $>$ the query intends) and verifies unit compatibility.
The mixing weight $\alpha_Q \in (0,1)$ is a \textit{learned} gate derived from the query representation.

While effective, the architecture of DeepQuant, illustrated in Figure~\ref{fig:method_overall}(a), 
introduces several practical challenges.
\textbf{(1) Pipeline dependence.}
The CQE module~\figlabeld{D-i} is mandatory for both queries and documents: quantity mentions are replaced with a placeholder token \texttt{[num]}, and extracted attributes (value, unit, and comparison operator) are attached as metadata.
Moreover, the disjoint scoring pipeline~\figlabeld{D-ii} requires separate pathways for text and quantity matching, which differs from standard MaxSim and prevents direct reuse of the standard ColBERT pipeline.
\textbf{(2) Error propagation.}
Although comparison operator uncertainty is marginalized, errors in \textit{quantity span detection} or \textit{unit normalization} within~\figlabeld{D-i} directly degrade $S_{\text{num}}$. DeepQuant's authors themselves acknowledge CQE as a single point of failure~\cite{agrawal-etal-2025-dense-deepquant}.
\textbf{(3) Score calibration.}
The learned gate $\alpha_Q$ aggregates into a single scalar per query; mis-estimation disproportionately shifts the balance between text and numerical scores, with limited robustness guarantees on out-of-domain queries or unseen unit expressions.
\textbf{(4) Inference overhead.}
The disjoint architecture~\figlabeld{D-ii} requires two separate embedding pathways to be executed and combined at retrieval time, increasing latency.

Crucially, DeepQuant's strategy is to \textit{isolate} the numerical score and \textit{add} it to the text score. In contrast, NumColBERT takes a complementary approach: it \textit{reshapes influence within the same MaxSim framework} so that numerical conditions can exert proportionally larger effects on the ranking decision without requiring architectural changes to the retrieval pipeline.

\subsection{Proposed Method: NumColBERT}

To overcome these limitations, we propose NumColBERT, which injects numerical reasoning capabilities directly into the dense embedding space while remaining non-intrusive at inference time and preserving full compatibility with the ColBERT inference pipeline.
As shown in Figure~\ref{fig:method_overall}(b), NumColBERT introduces a lightweight gating mechanism~\figlabeln{N-i} while keeping the retrieval scoring~\figlabeln{N-ii} identical to standard ColBERT.

Our key insight is that rather than computing a separate numerical score, we can \textit{redistribute influence} within the existing MaxSim aggregation so that numerical conditions contribute more decisively to the final ranking.
NumColBERT maintains the exact same scoring function as ColBERT:
\begin{equation}
    S(Q, D) = \sum_{\tilde{\mathbf{q}}_i \in \tilde{Q}} \max_{\mathbf{d}_j \in D} \tilde{\mathbf{q}}_i \cdot \mathbf{d}_j
\end{equation}
where $\tilde{\mathbf{q}}_i$ denotes the gated query embedding (defined below).
Crucially, document embeddings $\mathbf{d}_j$ are generated using a standard encoder, making the document index indistinguishable from a standard ColBERT index.
All innovations are applied solely to the query encoder and the training process.

NumColBERT introduces two lightweight modules on the query side, collectively shown as~\figlabeln{N-i} in Figure~\ref{fig:method_overall}(b): \textbf{Numerical Token Detector} and a \textbf{Numerical Gate}.
The Numerical Token Detector $P_\mathrm{num}: \mathbb{R}^d \to (0,1)$, where $d$ denotes the query embedding dimension, is a two-layer MLP with sigmoid activation that outputs a probability indicating whether a token is part of a numerical expression.
For a query token embedding $\mathbf{q}_i$, the gated embedding $\tilde{\mathbf{q}}_i$ is computed as:
\begin{equation}
    \tilde{\mathbf{q}}_i = \begin{cases} 
        g_i \cdot \mathbf{q}_i & \text{if } P_\mathrm{num}(\mathbf{q}_i) > \tau \\ 
        \mathbf{q}_i & \text{otherwise} 
    \end{cases}
\end{equation}
where $g_i = |Q| \cdot \sigma(\mathrm{MLP}(\mathbf{q}_i)) \in (0, |Q|)$ is a learned scalar gate computed via a two-layer MLP and a sigmoid function, and $\tau$ is a threshold hyperparameter (default $0.5$).

Intuitively, the gate learns to amplify tokens carrying critical numerical constraints (e.g., ``500'' in ``smartphones under 500 dollars'') while suppressing neutral numerical mentions (e.g., model numbers or dates).
The gate parameters $\mathbf{w}_g$ are trained end-to-end through the retrieval loss $\mathcal{L}_{\text{ret}}$, which backpropagates through the gated embeddings $\tilde{\mathbf{q}}_i$; no dedicated gate-specific loss is required.
Unlike prior gating approaches targeting efficiency through token pruning~\cite{hofstatter2022colberter} or general importance weighting~\cite{kang-etal-2025-trial}, our gate is specifically conditioned on numerical semantics, enabling fine-grained control over how numerical constraints influence the retrieval score.

At inference time, the retrieval scoring~\figlabeln{N-ii} operates identically to ColBERT: the document index remains a standard collection of token embeddings with no additional metadata, and the scoring function is the unmodified MaxSim.
On the query side, the gating mechanism~\figlabeln{N-i} applies two separate two-layer MLPs to each of the $|Q|$ query tokens. 
This introduces only $O(|Q|)$ additional computation, which is negligible compared to the encoder forward pass.
This design allows NumColBERT to leverage all existing ColBERT optimizations and infrastructure without modification.

\subsection{Training Objectives}

NumColBERT is trained with a multi-task learning framework that combines four complementary objectives.
The \textit{retrieval loss} $\mathcal{L}_{\text{ret}}$ optimizes the standard ColBERT ranking objective.
The \textit{numerical contrastive loss} $\mathcal{L}_{\text{cont}}$ explicitly shapes the embedding space so that documents satisfying numerical conditions are closer to the query than those violating them.
The \textit{detection loss} $\mathcal{L}_{\text{det}}$ trains the numerical token detector to accurately identify numerical expressions.
Finally, the \textit{numerical property prediction loss} $\mathcal{L}_{\text{prop}}$ regularizes the encoder by predicting numerical properties such as unit, magnitude, and comparison intent.
We describe each objective below.

\subsubsection{Retrieval Loss}
Following standard ColBERT, we use an in-batch cross-entropy loss based on MaxSim scores:
\begin{equation}
    \mathcal{L}_{\text{ret}} = -\frac{1}{B}\sum_{k=1}^{B} \log\frac{\exp(S(Q_k,D^+_k)/\tau_{\text{ret}})}{\sum_{D \in \mathcal{D}}\exp(S(Q_k,D)/\tau_{\text{ret}})}
\end{equation}
where $B$ is the batch size, $D^+_k$ denotes a positive (relevant) document for query $Q_k$, $\tau_{\text{ret}}$ is a temperature hyperparameter, and the denominator sums over all positives and negatives in the batch.
Given a mini-batch of triples $\{(Q_k, D_k^+, D_k^-)\}_{k=1}^B$, 
where $D^-_k$ denotes a negative (irrelevant) document for query $Q_k$,
we define the in-batch candidate set $\mathcal{D} = \{D_1^+, \ldots, D_B^+, D_1^-, \ldots, D_B^-\}$.

\subsubsection{Numerical Contrastive Loss}
To explicitly shape the embedding space for numerical reasoning, we introduce a multi-positive InfoNCE objective.
Let $\mathbf{q}_{\rm num}$ be the mean-pooled embedding of all tokens representing numerical values in query $Q$.
We define the numerical similarity as the maximum dot product between the query's mean-pooled numerical embedding and the document tokens:
\begin{equation}
    S_{\text{cont}}(Q, D) = \max_{\mathbf{d}_j \in D} \, \mathbf{q}_{\rm num} \cdot \mathbf{d}_j
\end{equation}
Note that this differs from the retrieval score $S(Q, D)$, which aggregates over \textit{all} query tokens; here we focus solely on tokens representing numerical values to directly supervise numerical matching.
For a query such as ``companies with revenue above 1 billion USD,'' the extracted numerical condition is represented as $(u, v, \mathit{cmp}) = (\mathrm{USD}, 10^9, >)$, and the token embeddings corresponding to the quantity mention ``1 billion USD'' are mean-pooled to form $\mathbf{q}_{\rm num}$.

Each query $Q$ contains a numerical condition with a comparison operator $\mathit{cmp}$, a unit $u$, and a value $v$.
Positive set $P$ is constructed based on two criteria: (1) \textit{numerical satisfaction}---whether the document's value satisfies the query's comparison condition, and (2) \textit{unit compatibility}---whether the document's unit matches the query's unit.
We define four strategies for constructing the positive set:
\begin{itemize}
    \item \textbf{Unit Only}: $P = \{D \in \mathcal{D} : u_D = u\}$, where $u_D$ denotes the unit of document $D$. This strategy considers only unit matching.
    \item \textbf{Numeric Only}: $P = \{D \in \mathcal{D} : v_D \; \mathit{cmp} \; v\}$, where $v_D$ is the numerical value in $D$. This strategy considers only whether the numerical condition is satisfied.
    \item \textbf{Joint}: $P = \{D \in \mathcal{D} : (u_D = u) \land (v_D \; \mathit{cmp} \; v)\}$. A document is positive only if both unit compatibility and numerical satisfaction hold.
    \item \textbf{Separate}: The unit and numeric criteria are applied as independent contrastive objectives, each with its own positive set.
\end{itemize}
We empirically compare these variants in our experiments.

Given a positive set $P_k$ for each query $Q_k$,
the numerical contrastive loss is defined as:
\begin{equation}
    \mathcal{L}_{\text{cont}}
    = -\frac{1}{|\mathcal{K}|}\sum_{k\in\mathcal{K}}
    \frac{1}{|P_k|}\sum_{D \in P_k}
    \log
    \frac{\exp\bigl(S_{\text{cont}}(Q_k, D)/\tau_{\text{cont}}\bigr)}
    {\sum\limits_{D' \in \mathcal{D}} \exp\bigl(S_{\text{cont}}(Q_k, D')/\tau_{\text{cont}}\bigr)}
\end{equation}
where $\mathcal{K} = \{k : |P_k| > 0\}$ denotes queries with at least one positive.

\subsubsection{Detection Loss}
The numerical token detector $P_{\mathrm{num}}$ is trained as a binary classification task using binary cross-entropy loss $\mathcal{L}_{\text{det}}$, where each token is labeled as a numerical or non-numerical value based on ground truth annotations.
These labels are obtained from the quantity annotations used to construct the training data: tokens aligned with the query-side quantity mention are labeled positive, and all remaining query tokens are labeled negative.

\subsubsection{Numerical Property Prediction Loss}
To enrich the numerical representations, we jointly train prediction heads (two-layer MLPs with/without sigmoid activation) 
on the mean-pooled numerical embedding ${\mathbf q}_{\rm num}$:
\begin{equation}
\mathcal{L}_{\text{prop}} = \mathcal{L}_{\text{unit}} + \mathcal{L}_{\text{mantissa}} + \mathcal{L}_{\text{exponent}} + \mathcal{L}_{\text{cond}}
\end{equation}
where $\mathcal{L}_{\text{unit}}$ is a cross-entropy loss for unit classification, $\mathcal{L}_{\text{mantissa}}$ and $\mathcal{L}_{\text{exponent}}$ are MSE losses for predicting the mantissa and exponent components in scientific notation, and $\mathcal{L}_{\text{cond}}$ is a cross-entropy loss for predicting the comparison operator ($=$, $>$, and $<$) from the numerical embedding ${\mathbf q}_{\rm num}$.
These auxiliary tasks act as regularizers, forcing the encoder to capture the unit, magnitude, and intent of numerical tokens.

\subsubsection{Composite Loss}
The final training objective combines all components with weighting coefficients:
\begin{equation}
    \mathcal{L} = \mathcal{L}_{\text{ret}} + \lambda_{\text{cont}}\,\mathcal{L}_{\text{cont}} + \lambda_{\text{det}}\,\mathcal{L}_{\text{det}} + \lambda_{\text{prop}}\,\mathcal{L}_{\text{prop}}
\end{equation}
where $\lambda_{\star}$ controls the relative importance of each objective.

\section{Experiments}

In this section, we evaluate the performance of NumColBERT. Specifically, we aim to answer the following research questions:

\begin{itemize}
    \item \textbf{RQ1 (Numerical Retrieval Performance)}: Does NumColBERT outperform standard ColBERT and achieve performance comparable to state-of-the-art retrieval models with numerical-specialized modules?
    \item \textbf{RQ2 (Ablation Studies)}: How do individual components, such as the Gating Mechanism and Numerical Contrastive Loss, contribute to the overall retrieval performance?
    \item \textbf{RQ3 (Generalization)}: Does NumColBERT maintain high performance on general document retrieval tasks (e.g., MS MARCO)?
    \item \textbf{RQ4 (Efficiency)}: Can NumColBERT leverage ColBERT's acceleration mechanisms to reduce query latency and index size while maintaining retrieval effectiveness?
    \item \textbf{RQ5 (Impact of Comparison Operators)}: How does NumColBERT's performance vary across different comparison operators, and how distinctly does the model represent these operators in its embedding space?
\end{itemize}

\subsection{Experimental Setup}

\subsubsection{Datasets}

We employ three datasets spanning different domains and evaluation objectives.
Table~\ref{tab:dataset_statistics} summarizes the statistics of our evaluation datasets.

\textbf{FinQuant} and \textbf{MedQuant} are quantity-aware sentence retrieval benchmarks introduced by Almasian et al.~\cite{almasian2024numbersmatterbringingquantityawareness-numbersmatter}.
Following their formulation, we treat each sentence as a retrieval unit.
FinQuant contains 306,291 sentences extracted from 473,375 news articles spanning economics, science, sports, and technology domains collected between 2018 and 2022.
MedQuant contains 153,252 sentences derived from the TREC Medical Records track~\cite{10.1145/2506583.2506624}.
Both benchmarks provide test queries with explicit numerical constraints and sentence-level relevance annotations.

For training on FinQuant and MedQuant, we adopt the data construction procedure of Almasian et al.~\cite{almasian2024numbersmatterbringingquantityawareness-numbersmatter}.
For each concept-unit pair, we generate equality, lower-bound, and upper-bound queries, and construct training triplets by pairing each query with sentences whose values satisfy the target constraint as positives and with sentences from the same concept-unit pair that violate the constraint as negatives.
We apply the same three augmentation operations as in Numbers Matter: \textit{concept expansion}, \textit{unit permutation}, and \textit{value permutation}.
Concept expansion replaces the concept with semantically related expressions, unit permutation varies the unit surface form, and value permutation rewrites the numeric value so as to change whether the sentence satisfies the target constraint.
Following Almasian et al., replacement values are sampled from values observed for the same concept-unit pair.
The test queries are disjoint from these generated training triplets.

\textbf{MS MARCO}~\cite{msmarco} is a large-scale passage retrieval benchmark widely used for evaluating general retrieval capabilities.
We evaluate on the MS MARCO passage reranking task, which contains 8.8 million passages extracted from web documents and approximately 500k real user queries from the Bing search engine.
Importantly, we did not filter these queries for quantity comparison intent, allowing us to assess whether our method maintains strong performance on general retrieval tasks without catastrophic forgetting.


Note that data augmentation was not applied for the MS MARCO experiments; the augmentation strategy was used exclusively for training models evaluated on FinQuant and MedQuant.

\begin{table}[t]
    \centering
    \caption{Dataset statistics. For FinQuant and MedQuant, ``Corpus'' denotes the number of sentences in the evaluation collection, and ``Train triplets'' denotes the number of generated training triplets used for fine-tuning.}
    \label{tab:dataset_statistics}
    \begin{tabular}{lccc}
    \toprule
    \textbf{Dataset} & \textbf{Corpus} & \textbf{Train triplets} & \textbf{Test queries} \\
    \midrule
    FinQuant & 306,291 & 2.4M & 420 \\
    MedQuant & 153,252 & 1.2M & 210 \\
    MS MARCO & 8.8M & 39.8M & 500K \\
    \bottomrule
\end{tabular}

\end{table}

For FinQuant and MedQuant, the ``Train triplets'' counts are derived from the generated fine-tuning sets used in our experiments, rather than from the benchmark statistics reported by Almasian et al.~\cite{almasian2024numbersmatterbringingquantityawareness-numbersmatter}.

\subsubsection{Baselines}
We compare NumColBERT against four groups of baselines.

\textit{Sparse Retrievers}: \textbf{BM25} is a classical lexical retrieval baseline based on term matching, and \textbf{SPLADE}~\cite{formal-etal-2021-splade} is a representative learned sparse retrieval model. SPLADE was fine-tuned from a pretrained checkpoint\footnote{\url{https://huggingface.co/naver/splade-cocondenser-ensembledistil}} on the target dataset (denoted as $\text{SPLADE}_{\text{ft}}$).
\textit{Dense Retriever}: \textbf{ColBERT}~\cite{santhanam-etal-2022-colbertv2} serves as the primary baseline to measure the improvement gained by our proposed method. We fine-tuned ColBERT from a pretrained checkpoint\footnote{\url{https://huggingface.co/colbert-ir/colbertv2.0}} on the target dataset (denoted as $\text{ColBERT}_{\text{ft}}$). For a fair comparison, $\text{ColBERT}_{\text{ft}}$ was trained using the same retrieval loss $\mathcal{L}_{\text{ret}}$, batch size, and temperature $\tau_{\text{ret}}$ as NumColBERT.
\textit{LLM-based Reranker}: We report \textbf{GPT-4o-mini} results to serve as a reference for the performance achievable by large-scale language models. Documents were reranked based on a 1-to-5 relevance score predicted by the model (see \cite{agrawal-etal-2025-dense-deepquant} for details).
\textit{Quantity-Aware Retrievers}: \textbf{QColBERT}~\cite{almasian2024numbersmatterbringingquantityawareness-numbersmatter} and \textbf{DeepQuant}~\cite{agrawal-etal-2025-dense-deepquant} are state-of-the-art numerical retrieval models.

\noindent\textbf{Note on Result Sources.}
For $\text{ColBERT}_{\text{ft}}$ and our proposed NumColBERT, we conducted experiments ourselves under identical training conditions.
Results for BM25, $\text{SPLADE}_{\text{ft}}$, and QColBERT (reported) were taken from Almasian et al.~\cite{almasian2024numbersmatterbringingquantityawareness-numbersmatter}.
We additionally reproduced QColBERT under the evaluation setup used in this paper and report these results as QColBERT (reproduced); all statistical tests against QColBERT use this reproduction.
DeepQuant and GPT-4o-mini results were sourced from Agrawal et al.~\cite{agrawal-etal-2025-dense-deepquant}, as the implementation of DeepQuant is not publicly released, making independent reproduction infeasible.

\begin{table*}[t]
    \centering
    \caption{Main results on FinQuant and MedQuant. \textbf{Bold} indicates the best result and \underline{underline} indicates the second best. QColBERT (reported) denotes the numbers reported by Almasian et al., while QColBERT (reproduced) denotes our reproduction under the evaluation setup used in this paper. Markers on NumColBERT denote significant improvements over the corresponding baseline in the same column under a Holm-corrected paired $t$-test ($p<0.05$): $^{\dagger}$ vs.\ $\mathrm{ColBERT}_{\mathrm{ft}}$ and $^{\ddagger}$ vs.\ QColBERT (reproduced).}
    \label{tab:main_benchmark}
    \newcolumntype{M}[1]{>{\centering\arraybackslash}p{#1}}
\begin{tabular}{@{}p{3.5cm}*{4}{M{1.2cm}}M{0.3cm}*{4}{M{1.2cm}}@{}}
\toprule
& \multicolumn{4}{c}{\textbf{FinQuant}} & & \multicolumn{4}{c}{\textbf{MedQuant}} \\
\cmidrule(lr){2-5} \cmidrule(lr){7-10}
\textbf{Model} & \small nDCG@10 & \small MRR@10 & \small P@10 & \small R@100 & & \small nDCG@10 & \small MRR@10 & \small P@10 & \small R@100 \\
\midrule
\multicolumn{10}{l}{\textit{Sparse Retrievers}} \\
\quad BM25  & 0.10 & 0.15 & 0.06 & 0.47 & & 0.08 & 0.11 & 0.05 & 0.37 \\
\quad $\mathrm{SPLADE}_{\mathrm{ft}}$ & 0.42 & 0.51 & 0.21 & 0.74 & & 0.29 & 0.37 & 0.15 & 0.63 \\
\midrule
\multicolumn{10}{l}{\textit{Dense Retriever}} \\
\quad $\mathrm{ColBERT}_{\mathrm{ft}}$ & 0.50 & 0.62 & 0.26 & 0.81 & & \underline{0.44} & 0.52 & \underline{0.22} & \underline{0.80} \\
\midrule
\multicolumn{10}{l}{\textit{LLM-based Reranker}} \\
\quad GPT-4o-mini & 0.36 & 0.52 & 0.17 & 0.69 & & 0.26 & 0.36 & 0.13 & 0.62 \\
\midrule
\multicolumn{10}{l}{\textit{Quantity-Aware Retrievers}} \\
\quad QBM25 & 0.41 & 0.53 & 0.21 & 0.55 & & 0.37 & 0.47 & 0.18 & 0.51 \\
\quad QSPLADE & 0.53 & 0.67 & 0.29 & 0.83 & & 0.38 & 0.52 & 0.19 & 0.70 \\
\quad QColBERT (reported) & \underline{0.56} & 0.69 & 0.30 & \underline{0.87} & & 0.37 & 0.51 & 0.18 & 0.73 \\
\quad QColBERT (reproduced) & 0.52 & 0.65 & 0.27 & 0.82 & & 0.36 & 0.49 & 0.17 & 0.68 \\
\quad DeepQuant & \textbf{0.59} & \textbf{0.73} & \textbf{0.32} & \textbf{0.88} & & \underline{0.44} & \underline{0.59} & 0.21 & \underline{0.80} \\
\midrule
\multicolumn{10}{l}{\textit{Ours}} \\
\quad \textbf{NumColBERT} & \textbf{0.59}$^{\dagger\ddagger}$ & \underline{0.71}$^{\dagger\ddagger}$ & \underline{0.31}$^{\dagger\ddagger}$ & \underline{0.87}$^{\dagger\ddagger}$ & & \textbf{0.50}$^{\dagger\ddagger}$ & \textbf{0.61}$^{\dagger\ddagger}$ & \textbf{0.25}$^{\dagger\ddagger}$ & \textbf{0.84}$^{\dagger\ddagger}$ \\
\bottomrule
\end{tabular}

\end{table*}

\subsubsection{Implementation Details}
NumColBERT was fine-tuned from the same pretrained checkpoint as $\text{ColBERT}_{\text{ft}}$ on the target dataset.
Training was conducted on a single NVIDIA H100 GPU for 5 epochs, which took approximately 18 hours.
We used the AdamW optimizer~\cite{LoshchilovH19} with a learning rate of $2 \times 10^{-5}$ and a linear warmup schedule over the first 10\% of training steps.
The batch size was set to 256 with gradient accumulation step of 1, and we applied gradient clipping with a maximum norm of 1.0.
All experiments used bfloat16 mixed-precision training with gradient checkpointing for memory efficiency.
For the loss function, the temperature parameters $\tau_{\text{ret}}$ and $\tau_{\text{cont}}$ were both set to 0.02.
The loss weights were set as follows: $\lambda_{\text{cont}} = 0.05$ for the numerical contrastive loss, $\lambda_{\text{det}} = 0.05$ for the detection loss, and $\lambda_{\text{prop}} = 0.05$ for the auxiliary losses (mantissa, exponent, and condition prediction).
These relatively small weights ensured that the primary retrieval loss $\mathcal{L}_{\text{ret}}$ remained the dominant component of the total loss, preventing the auxiliary objectives from overwhelming the core retrieval optimization.
The numerical detection threshold $\tau$ was set to 0.5.
Our implementation is publicly available\footnote{\url{https://github.com/fujimaki3968/NumColBERT}}.

\subsection{Main Results (RQ1)}

Table~\ref{tab:main_benchmark} reports FinQuant and MedQuant results; \textbf{bold} indicates the best result and \underline{underline} the second best.
NumColBERT achieves an nDCG@10 of 0.59 on FinQuant and 0.50 on MedQuant, substantially outperforming both the fine-tuned ColBERT baseline ($\mathrm{ColBERT}_{\mathrm{ft}}$; 0.50 and 0.44) and our reproduced QColBERT (0.52 and 0.36).
As marked by $^{\dagger}$ and $^{\ddagger}$ in Table~\ref{tab:main_benchmark}, these gains over both baselines are statistically significant on all four metrics for both datasets under a Holm-corrected paired $t$-test ($p<0.05$), with multiplicity controlled within each $(\text{dataset}, \text{metric})$ family of the three pairwise comparisons among NumColBERT, $\mathrm{ColBERT}_{\mathrm{ft}}$, and reproduced QColBERT; the margins over reproduced QColBERT are particularly large on MedQuant (e.g., $+0.14$ in nDCG@10 and $+0.16$ in R@100).
Importantly, NumColBERT matches the nDCG@10 performance of DeepQuant, despite employing a much simpler architecture that avoids the external CQE module for input preprocessing and the disjoint scoring pipeline that separately computes text and numerical scores required by DeepQuant. While NumColBERT shows slightly lower performance than DeepQuant on FinQuant in terms of MRR@10, P@10, and R@100, it achieves superior performance across all metrics on MedQuant. We attribute this to the nature of the MedQuant dataset, which likely requires a more balanced combination of text and numerical matching, as partly evidenced by the strong performance of the standard ColBERT model. As discussed later, NumColBERT retains robust general text matching capabilities, which appears to be a key factor in its superior performance on MedQuant. These results confirm that it is possible to achieve state-of-the-art numerical retrieval performance while preserving the standard ColBERT search pipeline.

\subsection{Ablation Studies (RQ2)}
To quantify the contribution of each proposed component in NumColBERT, we conducted ablation studies on FinQuant (Table~\ref{tab:ablation}).
The performance of the ablated models is statistically significantly lower than that of the original NumColBERT model (paired $t$-test with the Holm correction, $p < 0.05$). 

Removing the \textbf{Gating Mechanism} degrades performance significantly (0.592 $\to$ 0.545), underscoring its critical role in amplifying condition-relevant numerical tokens.
The exclusion of the \textbf{Numerical Contrastive Loss} ($\mathcal{L}_{\text{cont}}$) precipitates the sharpest performance drop (0.592 $\to$ 0.522), confirming that explicit embedding space shaping is fundamental for effective numerical reasoning.
Finally, while the \textbf{Numerical Property Prediction Loss} ($\mathcal{L}_{\text{prop}}$) offers more modest gains (0.592 $\to$ 0.581), it serves as a beneficial auxiliary objective, justifying its inclusion as a necessary component of NumColBERT.
Note that we do not report results for removing the Detection Loss ($\mathcal{L}_{\text{det}}$), 
as this loss is required to train $P_{\mathrm{num}}$, the core component of the Gating Mechanism. 
Thus, its removal is equivalent to the ``w/o Gating Mechanism'' configuration.

These results show that \textit{without our proposed components, standard ColBERT cannot achieve competitive numerical retrieval}.
The ablated models approach the $\mathrm{ColBERT}_{\mathrm{ft}}$ baseline (0.50), suggesting that high performance without our mechanisms would require complex architectures like DeepQuant's disjoint pipeline.

\begin{table}[t]
    \centering
    \small
    \caption{Ablation study on FinQuant. Each proposed component, Gating Mechanism, Numerical Contrastive Loss ($\mathcal{L}_{\text{cont}}$), and Numerical Property Prediction Loss ($\mathcal{L}_{\text{prop}}$), contributes to the final performance.}
    \label{tab:ablation}
    \begin{tabular}{@{}>{\raggedright\arraybackslash}p{4.4cm}cc@{}}
\toprule
Method & nDCG@10 & Recall@100 \\
\midrule
NumColBERT (Full) & 0.592 & 0.874 \\
\quad w/o Gating Mechanism & 0.545 & 0.830 \\
\quad w/o Num. Contrastive Loss ($\mathcal{L}_{\text{cont}}$) & 0.522 & 0.803 \\
\quad w/o Num. Prop. Pred. Loss ($\mathcal{L}_{\text{prop}}$) & 0.581 & 0.857 \\
\bottomrule
\end{tabular}

\end{table}

We also investigated the impact of different positive set strategies for the Numerical Contrastive Loss (Table~\ref{tab:contrastive_pattern}). Compared to NumColBERT without the Numerical Contrastive Loss, all positive set strategies achieve performance gains.
However, a Tukey HSD test ($\alpha = 0.05$) revealed a statistically significant difference solely between the ``w/o Numerical Contrastive Loss'' baseline and the ``Unit Only'' strategy, with no significant differences observed for any other pairs.
These results suggest that unit compatibility serves as a particularly effective supervision signal for the numerical contrastive loss objective.
Importantly, this does not imply that the model relies solely on unit matching; rather, fine-grained numerical comparison capabilities could be driven by the other losses (e.g., $\mathcal{L}_{\text{ret}}$, which optimizes the model to satisfy the full numerical condition (value, operator, and unit)).

\begin{table}[t]
    \centering
    \caption{Analysis of positive set strategies for Numerical Contrastive Loss. The first row shows the baseline without contrastive loss; all strategies yield substantial gains, with ``Unit Only'' performing best.}
    \label{tab:contrastive_pattern}
    \begin{tabular}{lcc}
\toprule
Positive Set Strategy & nDCG@10 & Recall@100 \\
\midrule
w/o Num. Contrastive Loss & 0.522 & 0.803 \\
\midrule
Joint & 0.581 & 0.866 \\
Separate & 0.590 & 0.866 \\
Numeric Only & 0.580 & 0.864 \\
Unit Only & 0.592 & 0.874 \\
\bottomrule
\end{tabular}
\end{table}

\subsection{Generalization (RQ3)}

A central question for numerically specialized retrieval is whether the model preserves effectiveness on general text retrieval tasks.
To assess this trade-off, we evaluate NumColBERT on the MS MARCO passage reranking task after joint training on both MS MARCO and FinQuant, where each training batch contains 50\% FinQuant samples and 50\% MS MARCO samples, while keeping the total number of epochs and computational cost identical to single-dataset training.

Table~\ref{tab:general_retrieval} reports MRR@10 for all methods to preserve direct comparability with DeepQuant, whose published generalization results are limited to this metric. To provide a less selective view of model behavior, we additionally report nDCG@10 for NumColBERT and the ColBERT baseline.
In the zero-shot setting (training only on FinQuant), NumColBERT achieves 0.268 MRR@10 on MS MARCO, compared to 0.210 for DeepQuant.
With joint training, NumColBERT improves to 0.356 MRR@10 on MS MARCO, recovering 95\% of the ColBERT baseline (0.376), while maintaining strong numerical retrieval effectiveness on FinQuant (0.708 MRR@10).
The same trend is reflected in nDCG@10: on MS MARCO, joint training improves NumColBERT from 0.321 to 0.415, which is close to the ColBERT baseline (0.436), whereas on FinQuant the corresponding change is small, from 0.592 to 0.586.
In contrast, DeepQuant reaches 0.332 MRR@10 on MS MARCO under joint training, retaining 88\% of the ColBERT baseline, although it remains slightly stronger on FinQuant MRR@10 (0.731 vs.\ 0.708).
Taken together, these results indicate that mixed training substantially recovers general-domain retrieval quality while largely preserving in-domain numerical retrieval performance.

\begin{table}[t]
    \centering
    \small
    \setlength{\tabcolsep}{3.5pt}
    \caption{General retrieval (MS MARCO) and numerical retrieval (FinQuant) performance. We report MRR@10 for all methods to preserve direct comparability with DeepQuant, and additionally report nDCG@10 where evaluation outputs are available. Here, \textit{zero} denotes the zero-shot setting trained only on FinQuant, and \textit{joint} denotes joint training on MS MARCO and FinQuant.}
    \label{tab:general_retrieval}
    \begin{tabular}{@{}lcccc@{}}
\toprule
& \multicolumn{2}{c}{\textbf{MS MARCO}} & \multicolumn{2}{c}{\textbf{FinQuant}} \\
\cmidrule(lr){2-3} \cmidrule(lr){4-5}
Method & MRR@10 & nDCG@10 & MRR@10 & nDCG@10 \\
\midrule
ColBERT & 0.376 & 0.436 & 0.373 & 0.303 \\
\midrule
DeepQuant (zero) & 0.210 & -- & 0.734 & -- \\
DeepQuant (joint) & 0.332 & -- & 0.731 & -- \\
NumColBERT (zero) & 0.268 & 0.321 & 0.710 & 0.592 \\
NumColBERT (joint) & 0.356 & 0.415 & 0.708 & 0.586 \\
\bottomrule
\end{tabular}

\end{table}





\subsection{Efficiency (RQ4)}

A key advantage of NumColBERT is that all architectural innovations are confined to the query encoder---the document index remains identical to standard ColBERT.
This design enables NumColBERT to leverage existing ColBERT-optimized retrieval engines without modification.
To empirically validate this compatibility, we evaluate both ColBERT and NumColBERT using \textbf{PLAID}~\cite{santhanam2022plaid}, a state-of-the-art retrieval engine that accelerates ColBERT search through centroid-based pruning and product quantization.
We demonstrate that PLAID accelerates NumColBERT as effectively as it does standard ColBERT.

\begin{table}[t]
    \centering
    \caption{Efficiency of NumColBERT / ColBERT with PLAID.  nbits=``-'' indicates the configuration without PLAID.}
    \label{tab:efficiency}
    \begin{tabular}{ccc}
\toprule
nbits & Latency (ms/query) & Index Size (GB) \\
\midrule
- & 2623 / 2580 & 5.1 / 5.1 \\
8 & 90 / 70 & 1.4 / 1.4 \\
4 & 85 / 71 & 0.75 / 0.75 \\
2 & 83 / 70 & 0.43 / 0.43 \\
1 & 78 / 66 & 0.27 / 0.27 \\
\bottomrule
\end{tabular}
\end{table}

\subsubsection{Experimental Settings}
All experiments were conducted on a server equipped with two Intel Xeon Gold 6226R CPUs (16 cores per socket, 2 threads per core, for a total of 64 threads) and one NVIDIA GeForce RTX 3090 GPU with 24 GB of high-bandwidth memory.
The GPU was used only for query encoding, while the retrieval process itself ran entirely on the CPU.
We focused on a key PLAID parameter that controls the trade-off between retrieval speed and accuracy: \textbf{nbits}, the number of bits used for product quantization.
Fewer bits reduce index size and accelerate scoring but introduce quantization error.
We systematically varied nbits from 1 to 8 and measured nDCG@10 on the FinQuant dataset for both ColBERT and NumColBERT, using identical PLAID settings.
The nprobe parameter, which controls the number of centroids probed during search, was fixed at 8 for all experiments.

\subsubsection{Results}

\begin{figure}[t]
    \centering
    \includegraphics[width=\linewidth]{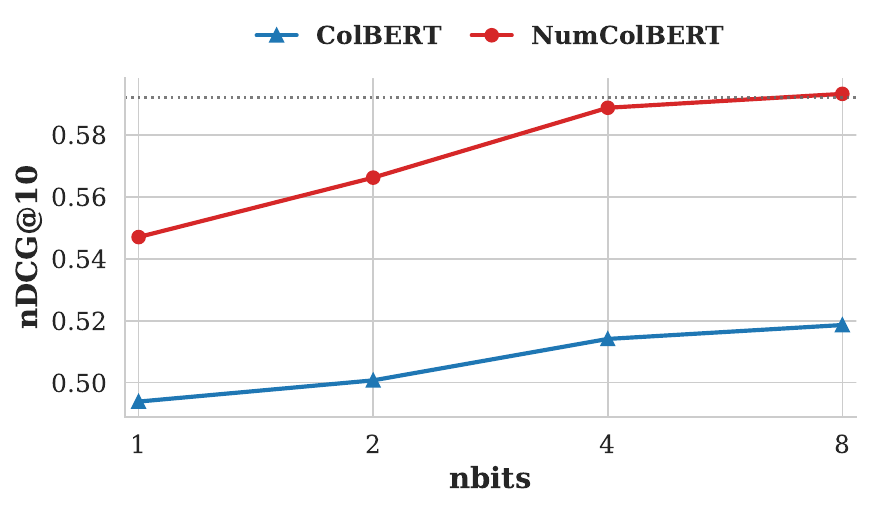}
    \caption{Performance degradation under PLAID quantization. Effect of varying nbits (quantization precision) on nDCG@10. Both ColBERT and NumColBERT exhibit similar degradation trends, suggesting that NumColBERT's document index maintains compatibility with PLAID.}
    \Description{Line plot showing nDCG@10 on FinQuant as a function of PLAID nbits for ColBERT and NumColBERT. Both models follow similar degradation curves as nbits decreases, with NumColBERT consistently above ColBERT.}
    \label{fig:plaid_efficiency}
\end{figure}

Table~\ref{tab:efficiency} reports query latency and index size across various nbits settings, 
while Figure~\ref{fig:plaid_efficiency} illustrates the performance degradation curves for both models as nbits decreases.
The results reveal two important findings:

\textbf{Efficient Query Processing and Compressed Index.}
As detailed in Table~\ref{tab:efficiency}, the integration of PLAID yields substantial gains in both query latency and storage efficiency for NumColBERT.
Notably, at nbits=8, where retrieval effectiveness remains on par with the configuration without PLAID, 
NumColBERT achieves a $>20\times$ speedup and a $>3\times$ reduction in index size compared to the configuration without PLAID.
Decreasing nbits further results in approximately proportional reductions in index size, accompanied by marginal decreases in query latency.
When compared to the standard ColBERT baseline, NumColBERT incurs a computational overhead of approximately 15--25\% in total query processing time, attributed to the additional computations required by the Gating Mechanism.

\textbf{Similar Degradation Trends.}
Figure~\ref{fig:plaid_efficiency} shows that at nbits=8, NumColBERT achieves an nDCG@10 of 0.59 on FinQuant, matching the results reported in Table~\ref{tab:main_benchmark}. This confirms that PLAID quantization introduces negligible performance loss at this setting.
As nbits decreases, both ColBERT and NumColBERT exhibit similar overall degradation trends.
While the general pattern is consistent, we observe that NumColBERT shows a slightly larger absolute performance drop compared to ColBERT. This may be attributed to its higher baseline performance, or to the finer-grained numerical representations being more sensitive to quantization error.
Nevertheless, this behavior suggests that NumColBERT's document representations remain structurally compatible with those of standard ColBERT, allowing the same quantization techniques to be applied with reasonable accuracy trade-offs.

These empirical results validate our design: by confining numerical awareness to the query encoder, NumColBERT preserves the standard ColBERT document index structure. This ensures compatibility with the existing PLAID infrastructure, allowing practitioners to deploy the model without modification and flexibly tune efficiency trade-offs. Furthermore, NumColBERT can inherit future ecosystem advancements for late-interaction retrieval models.


\begin{figure}[t]
    \centering
    \includegraphics[width=\linewidth]{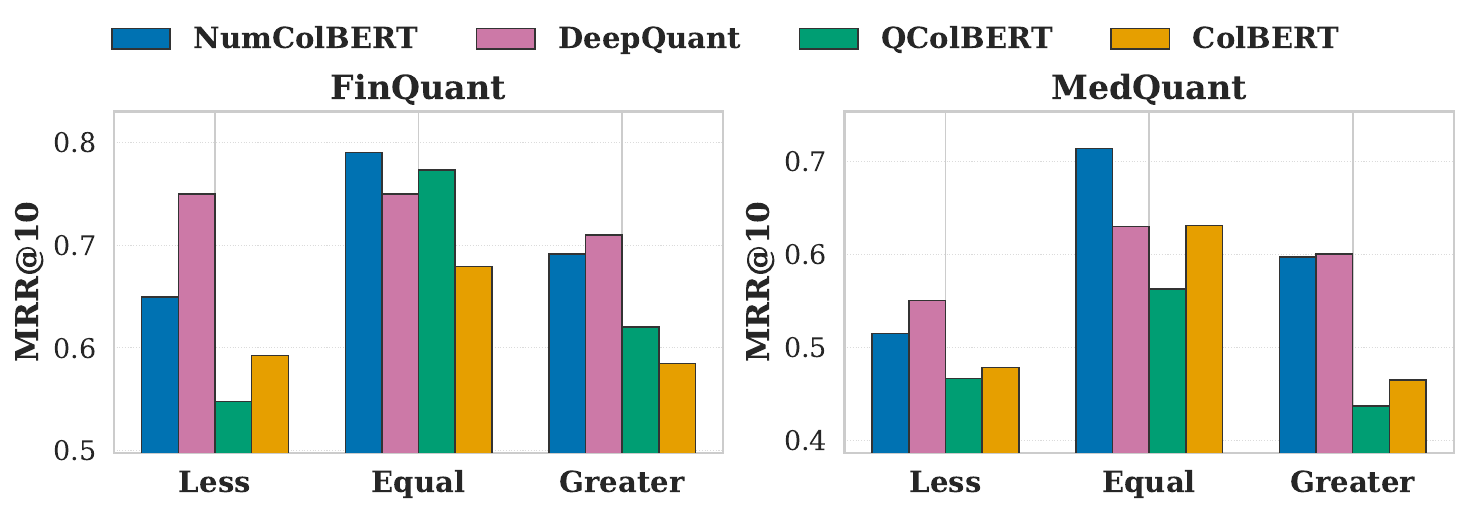}
    \caption{MRR@10 performance on FinQuant and MedQuant broken down by comparison operators (Less, Equal, and Greater). NumColBERT, DeepQuant, QColBERT (reproduced), and the fine-tuned ColBERT baseline are compared.}
    \Description{Bar chart of MRR@10 on FinQuant and MedQuant grouped by comparison operator (Less, Equal, Greater) for NumColBERT, DeepQuant, reproduced QColBERT, and fine-tuned ColBERT.}
    \label{fig:numq_analyze}
\end{figure}

\begin{figure}[t]
    \centering
    \includegraphics[width=\linewidth]{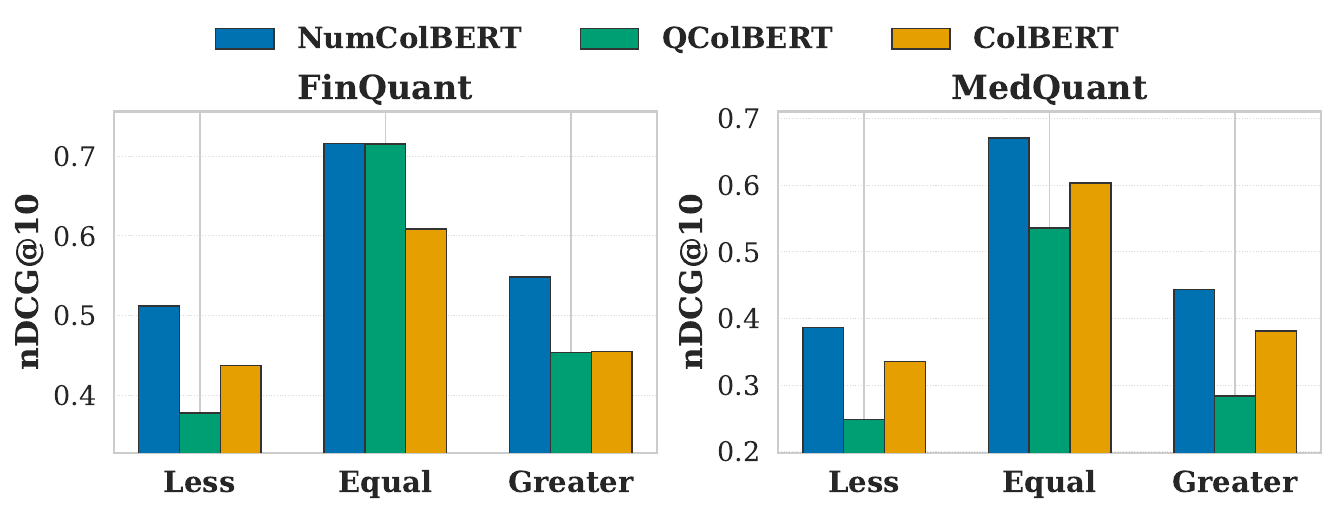}
    \caption{nDCG@10 performance by comparison operator on FinQuant and MedQuant for the reproducible models: NumColBERT, QColBERT (reproduced), and the fine-tuned ColBERT baseline.}
    \Description{Bar chart of nDCG@10 on FinQuant and MedQuant grouped by comparison operator (Less, Equal, Greater) for NumColBERT, reproduced QColBERT, and fine-tuned ColBERT.}
    \label{fig:numq_analyze_ndcg}
\end{figure}

\subsection{Impact of Comparison Operators (RQ5)}

\subsubsection{Performance by Comparison Operators}
Figure~\ref{fig:numq_analyze} breaks down performance by comparison operator on FinQuant and MedQuant: Equal ($=$), Greater ($>$), and Less ($<$).
We compare our proposed NumColBERT against DeepQuant, reproduced QColBERT, and the fine-tuned ColBERT baseline.
We report MRR@10 in Figure~\ref{fig:numq_analyze} to retain comparability with the condition-wise DeepQuant results reported by Agrawal et al.~\cite{agrawal-etal-2025-dense-deepquant}.
To verify that the observed pattern is not specific to MRR@10, Figure~\ref{fig:numq_analyze_ndcg} additionally reports nDCG@10 for the reproducible models.

On both FinQuant and MedQuant, NumColBERT achieves performance comparable to DeepQuant for the Equal and Greater conditions, while a performance gap remains for the Less condition, suggesting that modeling upper-bound constraints remains challenging.
NumColBERT also outperforms the fine-tuned ColBERT baseline across all comparison operators.
Against reproduced QColBERT, NumColBERT is strongest on the inequality conditions: in nDCG@10, it improves over QColBERT on all six condition-dataset pairs, with the largest margins appearing on Less and Greater queries, while FinQuant-Equal remains nearly tied (0.716 vs.\ 0.715).

On MedQuant, NumColBERT exhibits particularly strong performance on Equal conditions, surpassing all baseline methods including DeepQuant.
This may indicate that NumColBERT's unified architecture is better suited for capturing the contextual relationships between textual descriptors and numerical values.

\subsubsection{Embedding Analysis}
To gain deeper insights into how NumColBERT learns to represent numerical conditions, we visualize the numerical token embeddings using t-SNE~\cite{maaten2008visualizing}.
Specifically, we extract the embeddings of numerical tokens (identified by the detector $P_{\rm num}$) from queries in the MedQuant test set and project them to two dimensions.
Each point is colored according to its associated comparison operators: Equal ($=$), Greater ($>$), or Less ($<$).

\begin{figure}[t]
    \centering
    \includegraphics[width=\linewidth]{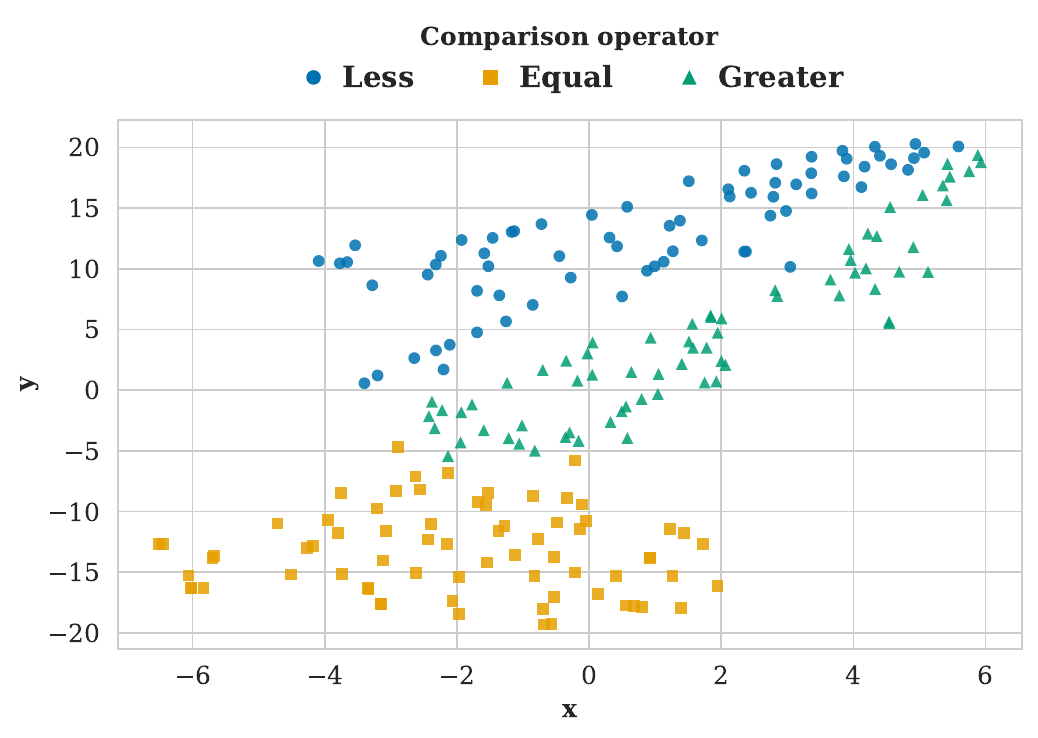}
    \caption{t-SNE visualization of NumColBERT's numerical token embeddings on MedQuant, colored by comparison operator (Less, Equal, Greater). Distinct clusters emerge for each comparison operator, suggesting that the model encodes numerical condition semantics rather than superficial value matching.}
    \Description{Two-dimensional t-SNE scatter plot of NumColBERT numerical token embeddings extracted from MedQuant queries, with points colored by comparison operator. Points associated with the same comparison operator form distinct, well-separated clusters.}
    \label{fig:tsne_embedding}
\end{figure}

Figure~\ref{fig:tsne_embedding} reveals that NumColBERT forms well-separated clusters aligned with comparison operators.
We used sklearn's t-SNE with 2 output dimensions, perplexity 30, learning rate \texttt{auto}, PCA initialization, and random seed 42. 
Embeddings associated with the same comparison operator tend to be grouped together, suggesting that the numerical condition serves as a strong signal in the learned embedding space.
This clustering behavior aligns with the design of our Numerical Contrastive Loss, which optimizes embeddings based on whether document values satisfy query conditions.
These findings provide interpretable evidence that NumColBERT learns to encode the semantics of numerical conditions, explaining its superior performance on numerically conditioned queries.

\section{Conclusion}

This paper presented NumColBERT, an inference-time non-intrusive method for numerically conditioned retrieval that remains compatible with the standard ColBERT pipeline.
Unlike prior approaches requiring external quantity extraction modules or separate scoring pathways, NumColBERT achieves numerical awareness through training-time innovations: the Numerical Gating Mechanism that amplifies condition-relevant numerical tokens within MaxSim, and the Numerical Contrastive Loss that shapes embeddings to reflect numerical satisfaction.

Experiments on FinQuant and MedQuant benchmarks demonstrated that NumColBERT substantially outperforms standard ColBERT fine-tuning and achieves accuracy competitive with specialized approaches such as DeepQuant.
Ablation studies confirmed the complementary contributions of the gating mechanism and contrastive learning, while generalization experiments showed relatively strong performance on general retrieval tasks.
Efficiency analyses verified that NumColBERT benefits from ColBERT optimizations such as PLAID, achieving acceleration comparable to that of standard ColBERT.

Our findings establish that numerically conditioned retrieval can be achieved with a non-intrusive inference pipeline, even though specialized numerical supervision remains necessary during training. This opens a promising direction for enhancing specialized reasoning capabilities through training objectives while preserving existing infrastructure advantages.

Future work includes extending NumColBERT to more complex numerical reasoning (e.g., multi-hop comparisons and arithmetic operations), improving less-than constraint handling, and exploring domain adaptation for specialized numerical conventions.

\begin{acks}
This work was supported by JSPS KAKENHI (JP24K03048 and JP23K28090) and JST PRESTO, Japan (JPMJPR25T2).

The authors acknowledge the peoples of the Woi Wurrung and Boon Wurrung language groups of the eastern Kulin Nation on whose unceded lands ACM SIGIR 2026 was hosted. We pay our respects to their Elders past and present, and extend that respect to all Aboriginal and Torres Strait Islander peoples today and their continuing connection to land, sea, sky, and community.
\end{acks}

\bibliographystyle{ACM-Reference-Format}
\bibliography{bibfile,software}

@inproceedings{10.1145/2506583.2506624,
author = {Voorhees, Ellen M.},
title = {The TREC Medical Records Track},
year = {2013},
isbn = {9781450324342},
publisher = {Association for Computing Machinery},
address = {New York, NY, USA},
url = {https://doi.org/10.1145/2506583.2506624},
doi = {10.1145/2506583.2506624},
booktitle = {Proceedings of the International Conference on Bioinformatics, Computational Biology and Biomedical Informatics},
pages = {239–246},
numpages = {8},
location = {Wshington DC, USA},
series = {BCB'13}
}

@inproceedings{li2021anasearch,
author = {Li, Tongliang and Fang, Lei and Lou, Jian-Guang and Li, Zhoujun and Zhang, Dongmei},
title = {AnaSearch: Extract, Retrieve and Visualize Structured Results from Unstructured Text for Analytical Queries},
year = {2021},
isbn = {9781450382977},
publisher = {Association for Computing Machinery},
address = {New York, NY, USA},
url = {https://doi.org/10.1145/3437963.3441694},
doi = {10.1145/3437963.3441694},
booktitle = {Proceedings of the 14th ACM International Conference on Web Search and Data Mining},
pages = {906–909},
numpages = {4},
location = {Virtual Event, Israel},
series = {WSDM '21}
}

@inproceedings{formal-etal-2021-splade,
  author = {Formal, Thibault and Piwowarski, Benjamin and Clinchant, St\'{e}phane},
  title = {SPLADE: Sparse Lexical and Expansion Model for First Stage Ranking},
  year = {2021},
  isbn = {9781450380379},
  publisher = {Association for Computing Machinery},
  address = {New York, NY, USA},
  url = {https://doi.org/10.1145/3404835.3463098},
  doi = {10.1145/3404835.3463098},
  booktitle = {Proceedings of the 44th International ACM SIGIR Conference on Research and Development in Information Retrieval},
  pages = {2288–2292},
  numpages = {5},
  location = {Virtual Event, Canada},
  series = {SIGIR '21}
}

@article{reddy2022shoppingqueriesdatasetlargescale,
  title={Shopping queries dataset: A large-scale {ESCI} benchmark for improving product search},
  author={Reddy, Chandan K and M{\`a}rquez, Llu{\'\i}s and Valero, Fran and Rao, Nikhil and Zaragoza, Hugo and Bandyopadhyay, Sambaran and Biswas, Arnab and Xing, Anlu and Subbian, Karthik},
  journal={arXiv preprint arXiv:2206.06588},
  year={2022}
}

@inproceedings{gill2025keywords,
  author = {Gill, Amritpal Singh and Patel, Sannikumar and Varga, P\'{e}ter and Miller, Patrick and Athanasiadis, Sakis},
  title = {From Keywords to Concepts: A Late Interaction Approach to Semantic Product Search on IKEA.com},
  year = {2025},
  isbn = {9798400715921},
  publisher = {Association for Computing Machinery},
  address = {New York, NY, USA},
  url = {https://doi.org/10.1145/3726302.3731948},
  doi = {10.1145/3726302.3731948},
  booktitle = {Proceedings of the 48th International ACM SIGIR Conference on Research and Development in Information Retrieval},
  pages = {4280–4283},
  numpages = {4},
  location = {Padua, Italy},
  series = {SIGIR '25}
}

@inproceedings{chen2022finqadatasetnumericalreasoning,
  title={FinQA: A Dataset of Numerical Reasoning over Financial Data},
    author = "Chen, Zhiyu  and
      Chen, Wenhu  and
      Smiley, Charese  and
      Shah, Sameena  and
      Borova, Iana  and
      Langdon, Dylan  and
      Moussa, Reema  and
      Beane, Matt  and
      Huang, Ting-Hao  and
      Routledge, Bryan  and
      Wang, William Yang",
  booktitle={EMNLP},
  pages={3697--3711},
  year={2021}
}

@article{jin2020diseasedoespatienthave,
  title={What disease does this patient have? a large-scale open domain question answering dataset from medical exams},
  author={Jin, Di and Pan, Eileen and Oufattole, Nassim and Weng, Wei-Hung and Fang, Hanyi and Szolovits, Peter},
  journal={Applied Sciences},
  volume={11},
  number={14},
  pages={6421},
  year={2021}
}

@inproceedings{devlin-etal-2019-bert,
    title = "{BERT}: Pre-training of Deep Bidirectional Transformers for Language Understanding",
    author = "Devlin, Jacob  and
      Chang, Ming-Wei  and
      Lee, Kenton  and
      Toutanova, Kristina",
    editor = "Burstein, Jill  and
      Doran, Christy  and
      Solorio, Thamar",
    booktitle = "Proceedings of the 2019 Conference of the North {A}merican Chapter of the Association for Computational Linguistics: Human Language Technologies, Volume 1 (Long and Short Papers)",
    month = jun,
    year = "2019",
    address = "Minneapolis, Minnesota",
    publisher = "Association for Computational Linguistics",
    url = "https://aclanthology.org/N19-1423",
    doi = "10.18653/v1/N19-1423",
    pages = "4171--4186",
}

@inproceedings{wallace-etal-2019-nlp,
    title = "Do {NLP} Models Know Numbers? Probing Numeracy in Embeddings",
    author = "Wallace, Eric  and
      Wang, Yizhong  and
      Li, Sujian  and
      Singh, Sameer  and
      Gardner, Matt",
    booktitle = "EMNLP-IJCNLP",
    year = "2019",
    pages = "5307--5315"
}

@inproceedings{mahendra-etal-2024-numbers,
    title = "Do Numbers Matter? Types and Prevalence of Numbers in Clinical Texts",
    author = "Mahendra, Rahmad  and
      Spina, Damiano  and
      Cavedon, Lawrence  and
      Verspoor, Karin",
    editor = "Demner-Fushman, Dina  and
      Ananiadou, Sophia  and
      Miwa, Makoto  and
      Roberts, Kirk  and
      Tsujii, Junichi",
    booktitle = "Proceedings of the 23rd Workshop on Biomedical Natural Language Processing",
    month = aug,
    year = "2024",
    address = "Bangkok, Thailand",
    publisher = "Association for Computational Linguistics",
    url = "https://aclanthology.org/2024.bionlp-1.32/",
    doi = "10.18653/v1/2024.bionlp-1.32",
    pages = "409--415"
}

@inproceedings{yang-etal-2025-numbercookbook,
  author       = {Haotong Yang and
                  Yi Hu and
                  Shijia Kang and
                  Zhouchen Lin and
                  Muhan Zhang},
  title        = {Number Cookbook: Number Understanding of Language Models and How to
                  Improve It},
  booktitle    = {The Thirteenth International Conference on Learning Representations,
                  {ICLR} 2025, Singapore, April 24-28, 2025},
  year         = {2025},
  bibsource    = {dblp computer science bibliography, https://dblp.org}
}

@inproceedings{ho2019qsearch,
author = {Ho, Vinh Thinh and Ibrahim, Yusra and Pal, Koninika and Berberich, Klaus and Weikum, Gerhard},
title = {Qsearch: Answering Quantity Queries from Text},
year = {2019},
isbn = {978-3-030-30792-9},
publisher = {Springer-Verlag},
address = {Berlin, Heidelberg},
url = {https://doi.org/10.1007/978-3-030-30793-6_14},
doi = {10.1007/978-3-030-30793-6_14},
booktitle = {The Semantic Web – ISWC 2019: 18th International Semantic Web Conference, Auckland, New Zealand, October 26–30, 2019, Proceedings, Part I},
pages = {237–257},
numpages = {21},
location = {Auckland, New Zealand}
}

@inproceedings{ho2020entities,
author = {Ho, Vinh Thinh and Pal, Koninika and Kleer, Niko and Berberich, Klaus and Weikum, Gerhard},
title = {Entities with Quantities: Extraction, Search, and Ranking},
year = {2020},
isbn = {9781450368223},
publisher = {Association for Computing Machinery},
address = {New York, NY, USA},
url = {https://doi.org/10.1145/3336191.3371860},
doi = {10.1145/3336191.3371860},
booktitle = {Proceedings of the 13th International Conference on Web Search and Data Mining},
pages = {833–836},
numpages = {4},
location = {Houston, TX, USA},
series = {WSDM '20}
}

@inproceedings{ho2021QuTE,
author = {Ho, Vinh Thinh and Pal, Koninika and Weikum, Gerhard},
title = {QuTE: Answering Quantity Queries from Web Tables},
year = {2021},
isbn = {9781450383431},
publisher = {Association for Computing Machinery},
address = {New York, NY, USA},
url = {https://doi.org/10.1145/3448016.3452763},
doi = {10.1145/3448016.3452763},
booktitle = {Proceedings of the 2021 International Conference on Management of Data},
pages = {2740–2744},
numpages = {5},
location = {Virtual Event, China},
series = {SIGMOD '21}
}

@inproceedings{ho2021extracting,
author = {Ho, Vinh Thinh and Pal, Koninika and Razniewski, Simon and Berberich, Klaus and Weikum, Gerhard},
title = {Extracting Contextualized Quantity Facts from Web Tables},
year = {2021},
isbn = {9781450383127},
publisher = {Association for Computing Machinery},
address = {New York, NY, USA},
url = {https://doi.org/10.1145/3442381.3450072},
doi = {10.1145/3442381.3450072},
booktitle = {Proceedings of the Web Conference 2021},
pages = {4033–4042},
numpages = {10},
location = {Ljubljana, Slovenia},
series = {WWW '21}
}

@inproceedings{ho2022enhancing,
author = {Ho, Vinh Thinh and Stepanova, Daria and Milchevski, Dragan and Str\"{o}tgen, Jannik and Weikum, Gerhard},
title = {Enhancing Knowledge Bases with Quantity Facts},
year = {2022},
isbn = {9781450390965},
publisher = {Association for Computing Machinery},
address = {New York, NY, USA},
url = {https://doi.org/10.1145/3485447.3511932},
doi = {10.1145/3485447.3511932},
booktitle = {Proceedings of the ACM Web Conference 2022},
pages = {893–901},
numpages = {9},
location = {Virtual Event, Lyon, France},
series = {WWW '22}
}

@article{roy2015reasoning,
    title = "Reasoning about Quantities in Natural Language",
    author = "Roy, Subhro  and
      Vieira, Tim  and
      Roth, Dan",
    editor = "Collins, Michael  and
      Lee, Lillian",
    journal = "Transactions of the Association for Computational Linguistics",
    volume = "3",
    year = "2015",
    address = "Cambridge, MA",
    publisher = "MIT Press",
    url = "https://aclanthology.org/Q15-1001/",
    doi = "10.1162/tacl_a_00118",
    pages = "1--13"
}

@inproceedings{ibrahim2016making,
author = {Ibrahim, Yusra and Riedewald, Mirek and Weikum, Gerhard},
title = {Making Sense of Entities and Quantities in Web Tables},
year = {2016},
isbn = {9781450340731},
publisher = {Association for Computing Machinery},
address = {New York, NY, USA},
url = {https://doi.org/10.1145/2983323.2983772},
doi = {10.1145/2983323.2983772},
booktitle = {Proceedings of the 25th ACM International on Conference on Information and Knowledge Management},
pages = {1703–1712},
numpages = {10},
location = {Indianapolis, Indiana, USA},
series = {CIKM '16}
}

@inproceedings{rybinski2023sciharvester,
author = {Rybinski, Maciej and Wan, Stephen and Karimi, Sarvnaz and Paris, Cecile and Jin, Brian and Huth, Neil and Thorburn, Peter and Holzworth, Dean},
title = {SciHarvester: Searching Scientific Documents for Numerical Values},
year = {2023},
isbn = {9781450394086},
publisher = {Association for Computing Machinery},
address = {New York, NY, USA},
url = {https://doi.org/10.1145/3539618.3591808},
doi = {10.1145/3539618.3591808},
booktitle = {Proceedings of the 46th International ACM SIGIR Conference on Research and Development in Information Retrieval},
pages = {3135–3139},
numpages = {5},
location = {Taipei, Taiwan},
series = {SIGIR '23}
}

@inproceedings{geva2020injecting,
    title = "Injecting Numerical Reasoning Skills into Language Models",
    author = "Geva, Mor  and
      Gupta, Ankit  and
      Berant, Jonathan",
    editor = "Jurafsky, Dan  and
      Chai, Joyce  and
      Schluter, Natalie  and
      Tetreault, Joel",
    booktitle = "Proceedings of the 58th Annual Meeting of the Association for Computational Linguistics",
    month = jul,
    year = "2020",
    address = "Online",
    publisher = "Association for Computational Linguistics",
    url = "https://aclanthology.org/2020.acl-main.89/",
    doi = "10.18653/v1/2020.acl-main.89",
    pages = "946--958"
}

@inproceedings{sundararaman2020methods,
    title = "Methods for Numeracy-Preserving Word Embeddings",
    author = "Sundararaman, Dhanasekar  and
      Si, Shijing  and
      Subramanian, Vivek  and
      Wang, Guoyin  and
      Hazarika, Devamanyu  and
      Carin, Lawrence",
    editor = "Webber, Bonnie  and
      Cohn, Trevor  and
      He, Yulan  and
      Liu, Yang",
    booktitle = "Proceedings of the 2020 Conference on Empirical Methods in Natural Language Processing (EMNLP)",
    month = nov,
    year = "2020",
    address = "Online",
    publisher = "Association for Computational Linguistics",
    url = "https://aclanthology.org/2020.emnlp-main.384/",
    doi = "10.18653/v1/2020.emnlp-main.384",
    pages = "4742--4753"
}

@inproceedings{thawani2021representing,
    title = "Representing Numbers in {NLP}: a Survey and a Vision",
    author = "Thawani, Avijit  and
      Pujara, Jay  and
      Ilievski, Filip  and
      Szekely, Pedro",
    editor = "Toutanova, Kristina  and
      Rumshisky, Anna  and
      Zettlemoyer, Luke  and
      Hakkani-Tur, Dilek  and
      Beltagy, Iz  and
      Bethard, Steven  and
      Cotterell, Ryan  and
      Chakraborty, Tanmoy  and
      Zhou, Yichao",
    booktitle = "Proceedings of the 2021 Conference of the North American Chapter of the Association for Computational Linguistics: Human Language Technologies",
    month = jun,
    year = "2021",
    address = "Online",
    publisher = "Association for Computational Linguistics",
    url = "https://aclanthology.org/2021.naacl-main.53/",
    doi = "10.18653/v1/2021.naacl-main.53",
    pages = "644--656"
}

@inproceedings{chen2023improving,
    title = "Improving Numeracy by Input Reframing and Quantitative Pre-Finetuning Task",
    author = "Chen, Chung-Chi  and
      Takamura, Hiroya  and
      Kobayashi, Ichiro  and
      Miyao, Yusuke",
    editor = "Vlachos, Andreas  and
      Augenstein, Isabelle",
    booktitle = "Findings of the Association for Computational Linguistics: EACL 2023",
    month = may,
    year = "2023",
    address = "Dubrovnik, Croatia",
    publisher = "Association for Computational Linguistics",
    url = "https://aclanthology.org/2023.findings-eacl.4/",
    doi = "10.18653/v1/2023.findings-eacl.4",
    pages = "69--77"
}

@inproceedings{sharma2024laying,
    title = "Laying Anchors: Semantically Priming Numerals in Language Modeling",
    author = "Sharma, Mandar  and
      Taware, Rutuja  and
      Koirala, Pravesh  and
      Muralidhar, Nikhil  and
      Ramakrishnan, Naren",
    editor = "Duh, Kevin  and
      Gomez, Helena  and
      Bethard, Steven",
    booktitle = "Findings of the Association for Computational Linguistics: NAACL 2024",
    month = jun,
    year = "2024",
    address = "Mexico City, Mexico",
    publisher = "Association for Computational Linguistics",
    url = "https://aclanthology.org/2024.findings-naacl.169/",
    doi = "10.18653/v1/2024.findings-naacl.169",
    pages = "2653--2660"
}

@inproceedings{sivakumar2025leverage,
    title = "How to Leverage Digit Embeddings to Represent Numbers?",
    author = "Sivakumar, Jasivan Alex  and
      Moosavi, Nafise Sadat",
    editor = "Rambow, Owen  and
      Wanner, Leo  and
      Apidianaki, Marianna  and
      Al-Khalifa, Hend  and
      Eugenio, Barbara Di  and
      Schockaert, Steven",
    booktitle = "Proceedings of the 31st International Conference on Computational Linguistics",
    month = jan,
    year = "2025",
    address = "Abu Dhabi, UAE",
    publisher = "Association for Computational Linguistics",
    url = "https://aclanthology.org/2025.coling-main.514/",
    pages = "7685--7697"
}

@inproceedings{hofstatter2020learning,
author = {Hofst\"{a}tter, Sebastian and Lipani, Aldo and Zlabinger, Markus and Hanbury, Allan},
title = {Learning to Re-Rank with Contextualized Stopwords},
year = {2020},
isbn = {9781450368599},
publisher = {Association for Computing Machinery},
address = {New York, NY, USA},
url = {https://doi.org/10.1145/3340531.3412079},
doi = {10.1145/3340531.3412079},
booktitle = {Proceedings of the 29th ACM International Conference on Information \& Knowledge Management},
pages = {2057–2060},
numpages = {4},
location = {Virtual Event, Ireland},
series = {CIKM '20}
}

@inproceedings{hofstatter2022colberter,
author = {Hofst\"{a}tter, Sebastian and Khattab, Omar and Althammer, Sophia and Sertkan, Mete and Hanbury, Allan},
title = {Introducing Neural Bag of Whole-Words with ColBERTer: Contextualized Late Interactions using Enhanced Reduction},
year = {2022},
isbn = {9781450392365},
publisher = {Association for Computing Machinery},
address = {New York, NY, USA},
url = {https://doi.org/10.1145/3511808.3557367},
doi = {10.1145/3511808.3557367},
booktitle = {Proceedings of the 31st ACM International Conference on Information \& Knowledge Management},
pages = {737–747},
numpages = {11},
location = {Atlanta, GA, USA},
series = {CIKM '22}
}

@inproceedings{kang-etal-2025-trial,
    title = "{TRIAL}: Token Relations and Importance Aware Late-interaction for Accurate Text Retrieval",
    author = "Kang, Hyukkyu  and
      Kim, Injung  and
      Han, Wook-Shin",
    editor = "Christodoulopoulos, Christos  and
      Chakraborty, Tanmoy  and
      Rose, Carolyn  and
      Peng, Violet",
    booktitle = "Proceedings of the 2025 Conference on Empirical Methods in Natural Language Processing",
    month = nov,
    year = "2025",
    address = "Suzhou, China",
    publisher = "Association for Computational Linguistics",
    url = "https://aclanthology.org/2025.emnlp-main.854/",
    doi = "10.18653/v1/2025.emnlp-main.854",
    pages = "16864--16877",
    ISBN = "979-8-89176-332-6"
}

@inproceedings{almasian-etal-2023-cqe,
    title = "{CQE}: A Comprehensive Quantity Extractor",
    author = {Almasian, Satya  and
      Kazakova, Vivian  and
      G{\"o}ldner, Philipp  and
      Gertz, Michael},
    editor = "Bouamor, Houda  and
      Pino, Juan  and
      Bali, Kalika",
    booktitle = "Proceedings of the 2023 Conference on Empirical Methods in Natural Language Processing",
    month = dec,
    year = "2023",
    address = "Singapore",
    publisher = "Association for Computational Linguistics",
    url = "https://aclanthology.org/2023.emnlp-main.793/",
    doi = "10.18653/v1/2023.emnlp-main.793",
    pages = "12845--12859"
}

@inproceedings{dua2019dropreadingcomprehensionbenchmark,
  title={DROP: A Reading Comprehension Benchmark Requiring Discrete Reasoning Over Paragraphs},
  author={Dua, Dheeru and Wang, Yizhong and Dasigi, Pradeep and Stanovsky, Gabriel and Singh, Sameer and Gardner, Matt},
  booktitle={NAACL-HLT},
  pages={2368--2378},
  year={2019}
}

@inproceedings{Haruki2025investigating-numqueries,
	author="Fujimaki, Haruki
	and P. Kato, Makoto",
	editor="Hauff, Claudia
	and Macdonald, Craig
	and Jannach, Dietmar
	and Kazai, Gabriella
	and Nardini, Franco Maria
	and Pinelli, Fabio
	and Silvestri, Fabrizio
	and Tonellotto, Nicola",
	title="Investigating the Performance of Dense Retrievers for Queries with Numerical Conditions",
	booktitle="Advances in Information Retrieval",
	year="2025",
	publisher="Springer Nature Switzerland",
	address="Cham",
	pages="210--218",
	isbn="978-3-031-88714-7"
}

@inproceedings{almasian2024numbersmatterbringingquantityawareness-numbersmatter,
    title = "Numbers Matter! Bringing Quantity-awareness to Retrieval Systems",
    author = "Almasian, Satya  and
      Bruseva, Milena  and
      Gertz, Michael",
    editor = "Al-Onaizan, Yaser  and
      Bansal, Mohit  and
      Chen, Yun-Nung",
    booktitle = "Findings of the Association for Computational Linguistics: EMNLP 2024",
    month = nov,
    year = "2024",
    address = "Miami, Florida, USA",
    publisher = "Association for Computational Linguistics",
    url = "https://aclanthology.org/2024.findings-emnlp.707/",
    doi = "10.18653/v1/2024.findings-emnlp.707",
    pages = "12120--12136"
}

@inproceedings{agrawal-etal-2025-dense-deepquant,
    title = "Dense Retrieval with Quantity Comparison Intent",
    author = "Agrawal, Prayas  and
      M, Nandeesh Kumar K  and
      Chelliah, Muthusamy  and
      Kumar, Surender  and
      Chakrabarti, Soumen",
    editor = "Che, Wanxiang  and
      Nabende, Joyce  and
      Shutova, Ekaterina  and
      Pilehvar, Mohammad Taher",
    booktitle = "Findings of the Association for Computational Linguistics: ACL 2025",
    month = jul,
    year = "2025",
    address = "Vienna, Austria",
    publisher = "Association for Computational Linguistics",
    url = "https://aclanthology.org/2025.findings-acl.1220/",
    doi = "10.18653/v1/2025.findings-acl.1220",
    pages = "23825--23839",
    ISBN = "979-8-89176-256-5"
}

@inproceedings{10.1145/3397271.3401075colbert,
    author = {Khattab, Omar and Zaharia, Matei},
    title = {ColBERT: Efficient and Effective Passage Search via Contextualized Late Interaction over BERT},
    year = {2020},
    isbn = {9781450380164},
    publisher = {Association for Computing Machinery},
    address = {New York, NY, USA},
    url = {https://doi.org/10.1145/3397271.3401075},
    doi = {10.1145/3397271.3401075},
    booktitle = {Proceedings of the 43rd International ACM SIGIR Conference on Research and Development in Information Retrieval},
    pages = {39–48},
    numpages = {10},
    location = {Virtual Event, China},
    series = {SIGIR '20}
}

@inproceedings{santhanam-etal-2022-colbertv2,
    title = "{C}ol{BERT}v2: Effective and Efficient Retrieval via Lightweight Late Interaction",
    author = "Santhanam, Keshav  and
      Khattab, Omar  and
      Saad-Falcon, Jon  and
      Potts, Christopher  and
      Zaharia, Matei",
    editor = "Carpuat, Marine  and
      de Marneffe, Marie-Catherine  and
      Meza Ruiz, Ivan Vladimir",
    booktitle = "Proceedings of the 2022 Conference of the North American Chapter of the Association for Computational Linguistics: Human Language Technologies",
    month = jul,
    year = "2022",
    address = "Seattle, United States",
    publisher = "Association for Computational Linguistics",
    url = "https://aclanthology.org/2022.naacl-main.272",
    doi = "10.18653/v1/2022.naacl-main.272",
    pages = "3715--3734",
}

@inproceedings{santhanam2022plaid,
    title = "{PLAID}: An Efficient Engine for Late Interaction Retrieval",
    author = "Santhanam, Keshav  and
      Khattab, Omar  and
      Potts, Christopher  and
      Zaharia, Matei",
    booktitle = "Proceedings of the 31st ACM International Conference on Information \& Knowledge Management",
    year = "2022",
    publisher = "Association for Computing Machinery",
    address = "New York, NY, USA",
    url = "https://doi.org/10.1145/3511808.3557325",
    doi = "10.1145/3511808.3557325",
    pages = "1747--1756",
}

@inproceedings{msmarco,
  author       = {Tri Nguyen and
                  Mir Rosenberg and
                  Xia Song and
                  Jianfeng Gao and
                  Saurabh Tiwary and
                  Rangan Majumder and
                  Li Deng},
  editor       = {Tarek Richard Besold and
                  Antoine Bordes and
                  Artur S. d'Avila Garcez and
                  Greg Wayne},
  title        = {{MS} {MARCO:} {A} Human Generated MAchine Reading COmprehension Dataset},
  booktitle    = {Proceedings of the Workshop on Cognitive Computation: Integrating
                  neural and symbolic approaches 2016 co-located with the 30th Annual
                  Conference on Neural Information Processing Systems {(NIPS} 2016),
                  Barcelona, Spain, December 9, 2016},
  series       = {{CEUR} Workshop Proceedings},
  volume       = {1773},
  publisher    = {CEUR-WS.org},
  year         = {2016},
  url          = {https://ceur-ws.org/Vol-1773/CoCoNIPS\_2016\_paper9.pdf},
  bibsource    = {dblp computer science bibliography, https://dblp.org}
}

@inproceedings{LoshchilovH19,
  title     = {Decoupled Weight Decay Regularization},
  author    = {Ilya Loshchilov and Frank Hutter},
  booktitle = {7th International Conference on Learning Representations, ICLR 2019, New Orleans, LA, USA, May 6--9, 2019},
  year      = {2019}
}

@article{maaten2008visualizing,
  title={Visualizing data using t-SNE},
  author={Maaten, Laurens van der and Hinton, Geoffrey},
  journal={Journal of machine learning research},
  volume={9},
  number={Nov},
  pages={2579--2605},
  year={2008}
}

@String{Computing = "Computing" }

@String{Computer = "{IEEE} Computer" }

@String{Springer = "Springer-Verlag" }

@ArtifactSoftware{R,
    title = {R: A Language and Environment for Statistical Computing},
    author = {{R Core Team}},
    organization = {R Foundation for Statistical Computing},
    address = {Vienna, Austria},
    year = {2019},
    url = {https://www.R-project.org/},
}


\end{document}